\documentstyle[aps,12pt]{revtex}

\begin{document}
\author{Xu Chang-Tan$^{1,2,3}$ and Liang J-Q$^1$}
\title{EPR spectrum via entangled states for an Exchange-Coupled Dimer of
Single-Molecule Magnets}
\address{$^1$Institute of Theoretical Physics, Shanxi University, Taiyuan 030006,\\
China\\
$^2$Department of Physics, Linyi Normal University, Shandong Linyi\\
276005,China\\
$^3$Department of Physics, Qufu Normal University, Shandong Qufu\\
273165,China}
\maketitle

\begin{abstract}
Multi-high-frequency electron paramagnetic resonance(EPR) spectrum for a
supermolecular dimer $\left[ Mn_4\right] _2$ of single-molecule magnets
recently reported [S. Hill, R. S. Edwards, N. Aliaga-Alcalde and G.
Christou(HEAC), Science 302, 1015 (2003)] is studied in terms of the
perturbation method in which the high-order corrections to the level
splittings of degenerate states are included. It is shown that the
corresponding eigenvectors are composed of entangled states of two
molecules. The EPR-peak positions are calculated in terms of the eigenstates
at various frequencies. From the best fit of theoretical level splittings
with the measured values we obtain the anisotropy constant and exchange
coupling which are in agreement with the corresponding values of
experimental observation. Our study confirms the prediction of HEAC that the
two $Mn_4$ units within the dimer are coupled quantum mechanically by the
antiferromagnetic exchange interaction and the supermolecular dimer
behaviors in analogy with artificially fabricated quantum dots.

\end{abstract}

\section{ Introduction}

Single-molecule magnets(SMM), which may be the smallest nanomagnets
exhibiting magnetization hysteresis-loop ( a classical property of
macroscopic magnets), straddle the interface between the characteristics of
classical and quantum worlds. The quantum tunneling of magnetization and
quantum phase interference known as macroscopic quantum effects have become
an attractive research field in recent years. The quantum effects in these
magnetic particles and clusters of nanometer size, such as $Mn_{12}(S=10),$Fe%
$_8(S=10)$ and $Mn_4(S=\frac 92)$ molecules, have been well studied$^{1-17}$
both experimentally and theoretically. Several proposals of possible quantum
computing schemes are also suggested using molecular magnets$^{18-20}$. The
high-frequency single crystal electron paramagnetic resonance(EPR) is a
powerful tool for the experimental study of various properties of
single-molecule magnets$^{21-24}$ . Wernsdorfer et al.$^{25}$ recently
pointed out that the supermolecular dimer $\left[ Mn_4\right] _2$ consisting
of two molecule magnets $Mn_4$ with antiferromagnetic exchange-coupling
exhibits a quite different quantum behavior from two individual $Mn_4$
molecules without coupling$^{25-28}$. It is therefore of great importance to
understand the effect of the exchange interaction which leads to energy
level splitting$^{27,31}$ and has been studied with the high-order
perturbation method$^{29}$. The step-like magnetization hysteresis-loop in a
supermolecular dimer have been demonstrated in terms of the numerical
solution of the time-dependent Schr\H{o}dinger equation$^{30}$. Recently the
multi-high-frequency EPR was used to probe the magnetic excitations of the
supermolecular dimer $\left[ Mn_4\right] _2$ by Hill et al$^{31}$. The
measured spectra well interpret both the quantum transitions involving
coherent superposition states of the two molecules and the phase decoherence
rate, which provide a compelling evidence that the molecules are coupled
quantum mechanically by the antiferromagnetic exchange interaction. In this
paper, we restudy the EPR transitions in the supermolecular dimer $\left[
Mn_4\right] _2$ employing the high-order perturbation method$^{29,32}$ and
obtain the high-order corrections of level splitting. In terms of the
corresponding eigenvectors accurate values of the anisotropy constant $D$
and the antiferromagnetic exchange-coupling $J$ are determined by the best
fit between the theoretical and measured level-splitting. The EPR- peak
positions are also calculated as a function of frequencies.

We in the following section first construct the explicit eigenvectors which
are seen to be composed of entangled states. It is then shown that the
ground states of the dimer are nearly (with probability $0.99883$) the
maximum entangled states, or Bell states which are useful in the quantum
computing. The main goal of this work is to display all eigenvalues and the
entanglements of the corresponding eigenstates in the dimer system. Our
theoretical studies shine more light on the prediction of Hill et al$^{31}$
that the supermolecular dimer with proper chemical design may be used as
practical quantum devices for quantum computing.

\section{Level Splitting and Eigenvectors}

Neglecting off-diagonal crystal field terms and inter-molecular
interactions, the effective spin Hamiltonian with a magnetic field applied
parallel to the easy axis of an isolate SMM has the form$^{31}$ $\hat{H}%
_i=-D_0\widehat{S}_{iz}^2-B_4^0\hat{O}_4^0+g_z\mu _BB_z\widehat{S}_{iz},$%
where $\widehat{S}_{iz}$ is the $z$-axis spin projection operator, and the
index $i$ $(=1,2)$ is used to label the two $Mn_4$ molecules in the dimer ; $%
D_0>$ $0$ denotes the uniaxial anisotropy constant; $B_4^0\hat{O}_4^0$
characterizes the fourth order uniaxial anisotropy where $\hat{O}_4^0=35%
\widehat{S}_z^4-[30S(S+1)-25]\widehat{S}_z^2-6S(S+1)+3S^2(S+1)^2$; $g$ is
the electronic $g$-factor; and $\mu _B$ is the Bohr magneton. The
Hamiltonian $\hat{H}_i$ may be written in the following form apart from a
trivial constant

\begin{equation}
\hat{H}_i=-D\widehat{S}_{iz}^2-B\widehat{S}_{iz}^4+g_z\mu _BB_z\widehat{S}%
_{iz},
\end{equation}
where $D=$ $D_0-30S(S+1)-25,$ $B=35B_4^0.$

For the supermolecular dimer with antiferromagnetic coupling the Hamiltonian
is seen to be

\begin{equation}
\hat{H}=\left[ \hat{H}_1+\hat{H}_2+J_z\hat{S}_{1z}\widehat{S}_{2z}\right] +%
\frac 12J_{xy}\left( \widehat{S}_{1+}\widehat{S}_{2-}+\widehat{S}_{1-}%
\widehat{S}_{2+}\right) ,
\end{equation}
where $\widehat{S}_{\pm }=$ $\widehat{S}_x\pm i$ $\widehat{S}_y$ with $J_z$
and $J_{xy}$ denoting the strengths of exchange coupling. In the following
we consider only the isotropic case that $J_z=$ $J_{xy}=J$ $>0$ which has
been verified experimentally$^{27}$. We furthermore assume that the exchange
coupling $J$ is much less than the anisotropy constant $D$. Then, $\hat{H}$
is rewritten as the perturbation form

\begin{equation}
\hat{H}=\hat{H}_0+\hat{V},
\end{equation}

\begin{equation}
\hat{H}_0=-D\left( \widehat{S}_{1z}^2+\widehat{S}_{2z}^2\right) -B\left( 
\widehat{S}_{1z}^4+\widehat{S}_{2z}^4\right) +J\widehat{S}_{1z}\widehat{S}%
_{2z}+g_z\mu _BH_z\left( \widehat{S}_{1z}+\widehat{S}_{2z}\right) ,
\end{equation}

\begin{equation}
\hat{V}=\frac 12J\left( \widehat{S}_{1+}\widehat{S}_{2-}+\widehat{S}_{1-}%
\widehat{S}_{2+}\right) ,
\end{equation}
where $\hat{H}_0$ is the zeroth-order Hamiltonian. The zeroth-order
eigenvectors of the dimer are direct products of the single-molecule
eigenvectors such that $\left| S_1,S_2,M_1,M_2\right\rangle $ or abbreviated
as$\left| M_1,M_2\right\rangle $, where $M_1$ and $M_2$ represent the
quantum numbers of spin operators $\hat{S}_{1z}$ and $\widehat{S}_{2z}$
respectively. The dimer has $(2S_1+1)(2S_2+1)$ energy levels which are
labeled by the spin quantum numbers $M_1$ and $M_2$ and given by

\begin{equation}
E_0=-D\left( M_1^2+M_2^2\right) -B\left( M_1^4+M_2^4\right) +JM_1M_2+g_z\mu
_BH_z\left( M_1+M_2\right) .
\end{equation}
It is seen obviously that the eigenstates $\left| M_1,M_2\right\rangle $ and 
$\left| M_2,M_1\right\rangle $ are degenerate since $%
E_0(M_1,M_2)=E_0(M_2,M_1)$. The two $Mn_4$ molecules in the $[Mn_4]_2$ dimer
are coupled by a weak exchange interaction via both the six C-H$\cdots $Cl
hydrogen bonds. Thus, we can treat $\hat{V}$ perturbationally.. The level
splitting of the pair of degenerate eigenstates $\left| M_1,M_2\right\rangle 
$ and $\left| M_2,M_1\right\rangle $ with $M_2>M_1$ appears only in the
chain of matrix elements connecting the states $\left|
M_1+k,M_2-k\right\rangle $ and $\left| M_1+k+1,M_2-k-1\right\rangle $ ,where 
$k=1,2,\cdots ,M_2-M_1-1$. Then, the level splitting becomes$^{29,32}$

\begin{eqnarray}
\Delta E_{M_1M_2,M_2M_1} &=&2\hat{V}_{M_1M_2,(M_1+1)(M_2-1)}\frac 1{%
E_{(M_1+1)(M_2-1)}-E_{M_1M_2}}\hat{V}_{(M_1+1)(M_2-1),(M_1+2)(M_2-2)} 
\nonumber \\
&&\times \frac 1{E_{(M_1+2)(M_2-2)}-E_{M_1M_2}}\cdots \hat{V}%
_{(M_2-1)(M_1+1),M_2M_1},
\end{eqnarray}
where 
\begin{equation}
\hat{V}_{M_1M_2,(M_1+1)(M_2-1)}=\left\langle M_1,M_2\right| \frac 12J\left( 
\widehat{S}_{1+}\widehat{S}_{2-}+\widehat{S}_{1-}\widehat{S}_{2+}\right)
\left| M_1+1,M_2-1\right\rangle =\frac J2h_{(M_1+1)(M_2-1)},
\end{equation}
and

\begin{equation}
h_{(M_1+1)(M_2-1)}=\left[ \left( S_1+M_1+1\right) \left( S_1-M_1\right)
\left( S_2-M_2+1\right) \left( S_2+M_2\right) \right] ^{\frac 12}.
\end{equation}
Since $S_1=S_2=S$ for the dimer case, we obtain the level splitting as 
\begin{equation}
\Delta E=\left\{ \prod_k\left[ 2D^{^{\prime }}-4Bk\left( M_2-M_1-k\right)
\right] \right\} ^{-1}\frac{J^{M_2-M_1}}{\left[ \left( M_1-M_2-1\right)
!\right] ^2}\frac{\left( S+M_2\right) !\left( S-M_1\right) !}{\left(
S-M_2\right) !\left( S+M_1\right) !},
\end{equation}
where 
\begin{equation}
D^{^{\prime }}=2D+J+4B\left( M_1+M_2\right) ^2-4BM_1M_2.
\end{equation}
Disregarding the fourth- power of anisotropy term in Eq..(4) i.e. $B=0,$ the
formula of level splitting can be simplified as

\begin{equation}
\Delta E=\left( 4D+2J\right) \left( \frac J{4D+2J}\right) ^{M_2-M_1}\frac{%
\left( S+M_2\right) !\left( S-M_1\right) !}{\left( S-M_2\right) !\left(
S+M_1\right) !}\left[ \frac 1{\left( M_2-M_1-1\right) !}\right] ^2.
\end{equation}
The state of the zeroth-order perturbation is obviously the superposition of
the states $\left| M_1,M_2\right\rangle $ and $\left| M_2,M_1\right\rangle $
such that

\begin{equation}
\left| \psi ^{\left( 0\right) }\right\rangle =a_{M_1M_2}^{(0)}\left|
M_1,M_2\right\rangle +a_{M_2M_1}^{(0)}\left| M_2,M_1\right\rangle .
\end{equation}
The first-order state is then obtained as

\begin{equation}
\left| \psi ^{\left( 1\right) }\right\rangle
=\sum_{k_1,k_2}a_{k_1k_2}^{\left( 1\right) }\left| k_1,k_2\right\rangle ,
\end{equation}
where 
\begin{equation}
a_{k_1k_2}^{\left( 1\right) }=\frac{a_{M_1M_2}^{(0)}\hat{V}%
_{k_1k_2,M_1M_2}+a_{M_2M_1}^{(0)}\hat{V}_{k_1k_2,M_2M_1}}{%
E_{M_2M_1}-E_{k_1k_2}},
\end{equation}
and $a_{M_1M_2}^{\left( 1\right) }=a_{M_2M_1}^{\left( 1\right) }=0.$ The $n$%
-th order state is

\begin{equation}
\left| \psi ^{\left( n\right) }\right\rangle
=\sum_{k_1,k_2}a_{k_1k_2}^{\left( n\right) }\left| k_1,k_2\right\rangle ,
\end{equation}
where

\begin{equation}
a_{k_1k_2}^{\left( n\right) }=\sum_{k_1^{^{\prime }},k_2^{^{\prime }}}\frac{%
\hat{V}_{k_1k_2,k_1^{^{\prime }}k_2^{^{\prime }}}}{E_{M_2M_1}-E_{k_1k_2}}%
a_{k_1^{^{\prime }}k_2^{^{\prime }}}^{\left( n-1\right) },
\end{equation}
and $(k_1,k_2)\neq (k_1^{^{\prime }},k_2^{^{\prime }})\neq \left(
M_1,M_2\right) \neq \left( M_2,M_1\right) ,$ $a_{M_1M_2}^{\left( m\right)
}=a_{M_2M_1}^{\left( m\right) }=0(m=1,2,\cdots ,n).$ For $n$ $%
=M_2-k_1=k_2-M_1\geq 1,$ we have

\begin{equation}
a_{k_1k_2}^{\left( n\right) }=\frac{\hat{V}_{k_1k_2,(k_1+1)(k_2-1)}}{%
E_{M_2M_1}-E_{k_1k_2}}\frac{\hat{V}_{(k_1+1)(k_2-1),(k_1+2)(k_2-2)}}{%
E_{M_2M_1}-E_{k_1+1,k_2-1}}\frac{\hat{V}_{(k_1+2)(k_2-2),(k_1+3)(k_2-3)}}{%
E_{M_2M_1}-E_{k_1+2,k_2-2}}\cdots \frac{\hat{V}_{(M_2-1)(M_1+1),M_2M_1}}{%
E_{M_2M_1}-E_{M_2-1,M_1+1}},
\end{equation}
and

\begin{eqnarray}
\hat{V}_{(k_1+l-1)(k_2-l+1),(k_1+l)(k_2-l)} &=&\frac J2\left\langle
k_1+l-1,k_2-l+1\right| \widehat{S}_{1-}\widehat{S}_{2+}\left|
k_1+l,k_2-l\right\rangle  \nonumber \\
&=&\frac J2\left[ \left( S-k_2+l\right) \left( S+k_2+1-l\right) \left(
S+k_1+l\right) \left( S-k_1+1-l\right) \right] ^{\frac 12},
\end{eqnarray}
where $l=1,2,\cdots ,n$. We finally obtain the coefficients $%
a_{k_1k_2}^{\left( n\right) }$ of the $n$-th order state

\begin{eqnarray}
a_{k_1k_2}^{\left( n\right) } &=&J^n\left\{ \prod_{l=0}^{n-1}\left[
M_{12}+l\left( k_1-k_2+l\right) \right] \left[ 2D^{^{\prime \prime
}}+4Bl\left( k_1-k_2+l\right) \right] \right\} ^{-1}  \nonumber \\
&&\times \left[ \frac{\left( S+M_2\right) !\left( S-M_1\right) !\left(
S-k_1\right) !\left( S+k_2\right) !}{\left( S-M_2\right) !\left(
S+M_1\right) !\left( S+k_1\right) !\left( S-k_2\right) !}\right] ^{\frac 12},
\end{eqnarray}
where $M_{12}=M_1M_2-k_1k_2$ , $D^{^{\prime \prime }}=2D+J+4B\left(
M_1+M_2\right) ^2-2B\left( M_1M_2+k_1k_2\right) .$ For the case $n^{^{\prime
}}=$ $k_1^{^{\prime }}-M_1=M_2-k_2^{^{\prime }}\ \geq 1,$we have

\begin{eqnarray}
a_{k_1^{^{\prime }}k_2^{^{\prime }}}^{(n^{^{\prime }})} &=&J^{n^{^{\prime
}}}\left\{ \prod_{l=0}^{n^{^{\prime }}-1}\left[ M_{12}^{^{\prime }}+l\left(
k_2^{^{\prime }}-k_1^{^{\prime }}+l\right) \right] \left[ 2D^{^{\prime
\prime \prime }}+4Bl\left( k_2^{^{\prime }}-k_1^{^{\prime }}+l\right)
\right] \right\} ^{-1}  \nonumber \\
&&\times \left[ \frac{\left( S+M_2\right) !\left( S-M_1\right) !\left(
S-k_2^{^{\prime }}\right) !\left( S+k_1^{^{\prime }}\right) !}{\left(
S-M_2\right) !\left( S+M_1\right) !\left( S+k_2^{^{\prime }}\right) !\left(
S-k_1^{^{\prime }}\right) !\ }\right] ^{\frac 12},
\end{eqnarray}
where $M_{12}^{^{\prime }}=M_1M_2-k_1^{^{\prime }}k_2^{^{\prime }}$ , $%
D^{^{\prime \prime \prime }}=2D+J+4B\left( M_1+M_2\right) ^2-2B\left(
M_1M_2+k_1^{^{\prime }}k_2^{^{\prime }}\right) .$

The zeroth-order perturbation wave functions for the non-degenerate states $%
\left| M_1,M_2\right\rangle $ with $M_1=M_2=M$ are seen to be $\left| \psi
^{\left( 0\right) }\right\rangle =2\left| M,M\right\rangle .$ The matrix
elements of operator $\hat{V}$ in Eq. 18 appear only in the chain connecting
the states $\left| k_1+l,k_2-l\right\rangle $ and $\left|
k_1+l+1,k_2-l-1\right\rangle .$ We can obtain the results by non-degenerate
perturbation theory as follows. When $a_{k_1k_2}^{\left( n\right) }\neq
0,a_{k_1k_2}^{\left( 1\right) }=a_{k_1k_2}^{\left( 2\right) }=\cdots
=a_{k_1k_2}^{\left( n-1\right) }=0,$ as a consequence eqs. $(16),(17),(20)$
and $(21)$ become the same as for the case of non-degenerate states.

Making use of eqs.$(13),(16),(20)$ and $(21),$ we can obtain all
eigenvectors for the quantum number range from $M=M_1+M_2=-9$ to $0$ , while
the eigenvectors for the quantum number range from $M=1$ to $9$ can be
obtained simply by the replacements: $M_1\rightarrow -M_1,M_2\rightarrow
-M_2 $ in the former case. The eigenstates are listed in appendix A (where $%
\left| M_1,M_2\right\rangle _S$ and $\left| M_1,M_2\right\rangle _A$ denote
the symmetric and antisymmetric states respectively ) with parameter values $%
D=0.72K$, $B=1.8\times 10^{-3}K,$ $J=0.10K$ which give rise to the best fit
between theoretical and measured level splitting. The parameters $D$\ and $B$%
\ determined here with the best fit are in agreement with the values
obtained by Hill et al$^{31}${\bf . }When $M_1\neq M_2,$ the normalized
symmetric $\left| M_1,M_2\right\rangle _S$ and the antisymmetric $\left|
M_1,M_2\right\rangle _A$ states are actually maximum entangled states with
the entanglement degree of von Neumann entropy that $E(\rho )=1$, where 
\[
E(\rho )=Tr\rho \log _2\rho , 
\]
and $\rho $ is the reduced density matrix$^{33}$. When $M_1=M_2,$ the states
are disentangled. Thus, the eigenvectors are, in general, composed of
maximum entangled states with definite probabilities. For example, $\left|
\psi _1\right\rangle _S$ is the disentangled state and$\left| \psi
_2\right\rangle _S,\left| \psi _2\right\rangle _A,\left| \psi
_3\right\rangle _A$ are maximum entangled states with $E(\rho )=1$. The
state $\left| \psi _5\right\rangle _A$ is composed of maximum entangled
states $\frac 1{\sqrt{2}}\left| -\frac 92,-\frac 32\right\rangle _A$ and $%
\frac 1{\sqrt{2}}\left| -\frac 72,-\frac 52\right\rangle _A$ with
probabilities $0.9445$ and $0.0555$ respectively. Moreover, the probability
of the maximum entangled states $\frac 1{\sqrt{2}}\left| -\frac 92,\frac 92%
\right\rangle _{S,A}$ which emerge in the states $\left| \psi
_{26}\right\rangle _{S,A}$ is $0.99883$ . It is also seen that the ground
states $\left| \psi _{26}\right\rangle _S$ and $\left| \psi
_{26}\right\rangle _A$ are very close to the maximum entangled states in a
small range of magnetic field values.

\section{EPR Transition}

In the EPR experiment, it has been seen that the exchange-coupling induced
single-spin transitions i.e. the transitions from $(M_1,M_2)$ to $%
(M_1+1,M_2) $ or from $(M_1,M_2)$ to $(M_1,M_2+1)$ can occur, however, the
single-spin transitions should also depend on the states of the other spin
within the dimer because of the exchange coupling. This dependence has
displayed its importance in the resonant macroscopic quantum tunneling too.%
{\bf \ }Most importantly the multi-high-frequency electron paramagnetic
resonance spectra of the dimer obtained by Wernsdorfer et al$^{27}$ and Hill
et al.$^{31}$ show a compelling evidence that the two molecules $Mn_4$ are
coupled quantum mechanically by the antiferromagnetic exchange-coupling
which leads to the energy level splitting.. In terms of Eq.(6) and Eq. (10),
we can calculate the energy level splitting caused by the perturbation part
of the Hamiltonian $\hat{V}.$ Thus, when the frequency $\nu $ is fixed to
the value $145GHz$ and the magnetic field $B_z$ varies quasi-statically to
reach a resonance point, the EPR transition matrix elements with the
selection rule $\Delta M=\pm 1$ can be accurately calculated using the level
splitting formula Eq. (10) with the help of the corresponding eigenvectors
shown in appendix A. The magnetic dipole perturbation allows transitions
only between states of the same symmetry. The positions of principal and
other possible EPR peaks are given in Table $1$ (resonance transitions
between higher energy levels are not listed ), and displayed schematically
in Fig. $1$. From Table $1$, we can find that principal resonance peaks
labeled by $(x)$, $(a)$ to $(i)$ are in agreement with the result of Hill et
al.$^{31}$ (see Fig. 2 and Fig. 3 in Ref. [31]). From the Fig. 3 of Ref.[31]
we can see that the resonance peaks labeled by $(l)$, $(m),$ $(n),(p)$ and $%
(q)$ with magnetic field values $B_z=2.75\pm 0.05,4.85\pm 0.05,2.20\pm
0.05,3.55\pm 0.05$ , $3.80\pm 0.05$ respectively, are also in agreement with
our result in Table $1$. We expect that the possible $EPR$ peaks calculated
at the higher field region for $B_z\geq 6$ tesla can be verified by future $%
EPR$ experiments. Moreover the relation between the field value $B_z$ and
frequency $\nu $ can be obtained from the resonance condition 
\begin{equation}
h\nu =E\left( M+1\right) -E\left( M\right) =\Delta E+g_z\mu _BB_z,
\end{equation}
where the energy $E\left( M\right) $ can be accurately calculated from the
level splitting formula Eq. (10) and the eigenvectors in appendix A. The
level splitting{\bf \ }$\Delta E$ is shown in Table $2$ for various $EPR$
transitions. The calculated positions of $EPR$ peaks depending on
frequencies are shown in Fig. $2.$

\section{Conclusion}

The EPR spectra in the supermolecular dimer $\left[ Mn_4\right] _2$ with the
antiferromagnetic exchange-coupling between two molecule magnets are studied
in terms of the high-order perturbation method which allows us to obtain the
level splitting of the degenerate pairing states and the corresponding
eigenvectors. It is shown that the eigenvectors are composed of maximum
entangled states with definite probabilities. This observation shines more
light on the potential application of molecular magnets in quantum
information. The theoretical results are in good agreement with the measured
values$^{31}$ . The EPR peaks are obtained in terms of the high-order level
splitting formula and eigenvectors from which the value of exchange coupling 
$J=0.10K$ is determined and is seen in accord with the result of the
resonance quantum tunneling$^{25-29}$. Both experimental and theoretical
investigations of EPR for the supermolecular dimers are of great importance
for a better understanding the dynamics of such systems which are potential
devices for quantum computing.

{\bf ACKNOWLEDGMENT}

This work was supported by the Natural Science Foundation of China under
Grant No. 10475053.

{\bf Appendix A: Eigenvectors} 
\[
\left| \psi _1\right\rangle _S=\left| -\frac 92,-\frac 92\right\rangle , 
\]

\[
\left| \psi _2\right\rangle _{S,A}=\frac 1{\sqrt{2}}\left| -\frac 92,-\frac 7%
2\right\rangle _{S,A}=\frac 1{\sqrt{2}}\left( \left| -\frac 92,-\frac 72%
\right\rangle \pm \left| -\frac 72,-\frac 92\right\rangle \right) , 
\]

\[
\left| \psi _3\right\rangle _A=\frac 1{\sqrt{2}}\left| -\frac 92,-\frac 52%
\right\rangle _A, 
\]

\[
\left| \psi _3\right\rangle _S=0.64013\left| -\frac 92,-\frac 52%
\right\rangle _S-0.42482\left| -\frac 72,-\frac 72\right\rangle , 
\]

\[
\left| \psi _4\right\rangle _S=0.97356\left| -\frac 72,-\frac 72%
\right\rangle +0.16152\left| -\frac 92,-\frac 52\right\rangle _S, 
\]

\[
\left| \psi _5\right\rangle _S=0.69909\left| -\frac 92,-\frac 32%
\right\rangle _S-0.1062\left| -\frac 72,-\frac 52\right\rangle _S, 
\]

\[
\left| \psi _5\right\rangle _A=0.68721\left| -\frac 92,-\frac 32%
\right\rangle _A-0.16656\left| -\frac 72,-\frac 52\right\rangle _A, 
\]

\[
\left| \psi _6\right\rangle _{S,A}=0.69375\left| -\frac 72,-\frac 52%
\right\rangle _{S,A}+0.13677\left| -\frac 92,-\frac 32\right\rangle _{S,A}, 
\]

\[
\left| \psi _7\right\rangle _S=0.6992\left| -\frac 92,-\frac 12\right\rangle
_S-0.10364\left| -\frac 72,-\frac 32\right\rangle _S+0.02744\left| -\frac 52%
,-\frac 52\right\rangle , 
\]

\[
\left| \psi _7\right\rangle _A=0.70017\left| -\frac 92,-\frac 12%
\right\rangle _A-0.09882\left| -\frac 72,-\frac 32\right\rangle _A, 
\]

\[
\left| \psi _8\right\rangle _S=0.55604\left| -\frac 72,-\frac 32%
\right\rangle _S+0.08045\left| -\frac 92,-\frac 12\right\rangle
_S-0.6072\left| -\frac 52,-\frac 52\right\rangle , 
\]

\[
\left| \psi _8\right\rangle _A=0.69982\left| -\frac 72,-\frac 32%
\right\rangle _A+0.10125\left| -\frac 92,-\frac 12\right\rangle _A, 
\]

\[
\left| \psi _9\right\rangle _S=0.93217\left| -\frac 52,-\frac 52%
\right\rangle +0.25448\left| -\frac 72,-\frac 32\right\rangle
_S+0.02767\left| -\frac 92,-\frac 12\right\rangle _S, 
\]

\[
\left| \psi _{10}\right\rangle _S=0.7026\left| -\frac 92,\frac 12%
\right\rangle _S-0.07938\left| -\frac 72,-\frac 12\right\rangle
_S+0.00728\left| -\frac 52,-\frac 32\right\rangle _S, 
\]

\[
\left| \psi _{10}\right\rangle _A=0.70256\left| -\frac 92,\frac 12%
\right\rangle _A-0.07964\left| -\frac 72,-\frac 12\right\rangle
_A+0.00846\left| -\frac 52,-\frac 32\right\rangle _A, 
\]

\[
\left| \psi _{11}\right\rangle _S=0.6887\left| -\frac 72,-\frac 12%
\right\rangle _S+0.07793\left| -\frac 92,\frac 12\right\rangle
_S-0.14008\left| -\frac 52,-\frac 32\right\rangle _S, 
\]

\[
\left| \psi _{11}\right\rangle _A=0.65389\left| -\frac 72,-\frac 12%
\right\rangle _A+0.074\left| -\frac 92,\frac 12\right\rangle
_A-0.25876\left| -\frac 52,-\frac 32\right\rangle _A, 
\]

\[
\left| \psi _{12}\right\rangle _{S,A}=0.67168\left| -\frac 52,-\frac 32%
\right\rangle _{S,A}+0.20121\left| -\frac 72,-\frac 12\right\rangle
_{S,A}+0.09147\left| -\frac 92,\frac 12\right\rangle _{S,A}, 
\]

\[
\left| \psi _{13}\right\rangle _S=0.70425\left| -\frac 92,\frac 32%
\right\rangle _S-0.0633\left| -\frac 72,\frac 12\right\rangle
_S+0.0049\left| -\frac 52,-\frac 12\right\rangle _S-7.49\times 10^{-4}\left|
-\frac 32,-\frac 32\right\rangle , 
\]

\[
\left| \psi _{13}\right\rangle _A=0.70425\left| -\frac 92,\frac 32%
\right\rangle _A-0.06328\left| -\frac 72,\frac 12\right\rangle
_A+0.00484\left| -\frac 52,-\frac 12\right\rangle _A, 
\]

\[
\left| \psi _{14}\right\rangle _S=0.68766\left| -\frac 72,\frac 12%
\right\rangle _S+0.0618\left| -\frac 92,\frac 32\right\rangle
_S-0.14849\left| -\frac 52,-\frac 12\right\rangle _S+0.04995\left| -\frac 32%
,-\frac 32\right\rangle , 
\]

\[
\left| \psi _{14}\right\rangle _A=0.69083\left| -\frac 72,\frac 12%
\right\rangle _A+0.06208\left| -\frac 92,\frac 32\right\rangle
_A-0.13749\left| -\frac 52,-\frac 12\right\rangle _A, 
\]

\[
\left| \psi _{15}\right\rangle _S=0.49541\left| -\frac 52,-\frac 12%
\right\rangle _S+0.10279\left| -\frac 72,\frac 12\right\rangle
_S+0.00581\left| -\frac 92,\frac 32\right\rangle _S-0.69853\left| -\frac 32,-%
\frac 32\right\rangle , 
\]

\[
\left| \psi _{15}\right\rangle _A=0.69232\left| -\frac 52,-\frac 12%
\right\rangle _A+0.14364\left| -\frac 72,\frac 12\right\rangle
_A+0.00812\left| -\frac 92,\frac 32\right\rangle _A, 
\]

\[
\left| \psi _{16}\right\rangle _S=0.8928\left| -\frac 32,-\frac 32%
\right\rangle +0.31471\left| -\frac 52,-\frac 12\right\rangle
_S+0.04908\left| -\frac 72,\frac 12\right\rangle _S+0.00247\left| -\frac 92,%
\frac 32\right\rangle _S, 
\]

\[
\left| \psi _{17}\right\rangle _S=0.70536\left| -\frac 92,\frac 52%
\right\rangle _S-0.04963\left| -\frac 72,\frac 32\right\rangle
_S+3.012\times 10^{-3}\left| -\frac 52,\frac 12\right\rangle _S-1.68\times
10^{-4}\left| -\frac 32,-\frac 12\right\rangle _S, 
\]

\[
\left| \psi _{17}\right\rangle _A=0.70536\left| -\frac 92,\frac 52%
\right\rangle _A-0.04963\left| -\frac 72,\frac 32\right\rangle
_A+3.012\times 10^{-3}\left| -\frac 52,\frac 12\right\rangle
_A-0.00019\left| -\frac 32,-\frac 12\right\rangle _A, 
\]

\[
\left| \psi _{18}\right\rangle _S=0.69716\left| -\frac 72,\frac 32%
\right\rangle _S+0.04905\left| -\frac 92,\frac 52\right\rangle
_S-0.10696\left| -\frac 52,\frac 12\right\rangle _S+0.01129\left| -\frac 32,-%
\frac 12\right\rangle _S, 
\]

\[
\left| \psi _{18}\right\rangle _A=0.69696\left| -\frac 72,\frac 32%
\right\rangle _A+0.04904\left| -\frac 92,\frac 52\right\rangle
_A-0.10782\left| -\frac 52,\frac 12\right\rangle _A+0.01455\left| -\frac 32,-%
\frac 12\right\rangle _A, 
\]

\[
\left| \psi _{19}\right\rangle _S=0.68215\left| -\frac 52,\frac 12%
\right\rangle _S+0.10494\left| -\frac 72,\frac 32\right\rangle
_S+0.00447\left| -\frac 92,\frac 52\right\rangle _S-0.15375\left| -\frac 32,-%
\frac 12\right\rangle _S, 
\]

\[
\left| \psi _{19}\right\rangle _A=0.62554\left| -\frac 52,\frac 12%
\right\rangle _A+0.09623\left| -\frac 72,\frac 32\right\rangle
_A+0.0041\left| -\frac 92,\frac 52\right\rangle _A-0.3153\left| -\frac 32,-%
\frac 12\right\rangle _A, 
\]

\[
\left| \psi _{20}\right\rangle _{S,A}=0.66389\left| -\frac 32,-\frac 12%
\right\rangle _{S,A}+0.24213\left| -\frac 52,\frac 12\right\rangle
_{S,A}+0.02495\left| -\frac 72,\frac 32\right\rangle _{S,A}+0.00089\left| -%
\frac 92,\frac 52\right\rangle _{S,A}, 
\]

\begin{eqnarray*}
\left| \psi _{21}\right\rangle _S &=&0.70614\left| -\frac 92,\frac 72%
\right\rangle _S-0.03701\left| -\frac 72,\frac 52\right\rangle _S+1.75\times
10^{-3}\left| -\frac 52,\frac 32\right\rangle _S \\
&&-8.163\times 10^{-5}\left| -\frac 32,\frac 12\right\rangle _S+7.77\times
10^{-6}\left| -\frac 12,-\frac 12\right\rangle ,
\end{eqnarray*}

\[
\left| \psi _{21}\right\rangle _A=0.70614\left| -\frac 92,\frac 72%
\right\rangle _A-0.03701\left| -\frac 72,\frac 52\right\rangle _S+1.75\times
10^{-3}\left| -\frac 52,\frac 32\right\rangle _A-8.163\times 10^{-5}\left| -%
\frac 32,\frac 12\right\rangle _A, 
\]

\begin{eqnarray*}
\left| \psi _{22}\right\rangle _S &=&0.70148\left| -\frac 72,\frac 52%
\right\rangle _S+0.03676\left| -\frac 92,\frac 72\right\rangle
_S-0.08077\left| -\frac 52,\frac 32\right\rangle _S \\
&&+0.00723\left| -\frac 32,\frac 12\right\rangle _S-1.236\times
10^{-3}\left| -\frac 12,-\frac 12\right\rangle ,
\end{eqnarray*}

\[
\left| \psi _{22}\right\rangle _A=0.70148\left| -\frac 72,\frac 52%
\right\rangle _S+0.03676\left| -\frac 92,\frac 72\right\rangle
_A-0.08075\left| -\frac 52,\frac 32\right\rangle _A+0.0071\left| -\frac 32,%
\frac 12\right\rangle _A, 
\]

\[
\left| \psi _{23}\right\rangle _S=0.6801\left| -\frac 52,\frac 32%
\right\rangle _S+0.0783\left| -\frac 72,\frac 52\right\rangle
_S+0.00242\left| -\frac 92,\frac 72\right\rangle _S-0.17111\left| -\frac 32,%
\frac 12\right\rangle _S+0.0639\left| -\frac 12,-\frac 12\right\rangle , 
\]

\[
\left| \psi _{23}\right\rangle _A=0.68525\left| -\frac 52,\frac 32%
\right\rangle _A+0.07889\left| -\frac 72,\frac 52\right\rangle
_S+0.00244\left| -\frac 92,\frac 72\right\rangle _A-0.15559\left| -\frac 32,%
\frac 12\right\rangle _A, 
\]

\begin{eqnarray*}
\left| \psi _{24}\right\rangle _S &=&0.46545\left| -\frac 32,\frac 12%
\right\rangle _S+0.1114\left| -\frac 52,\frac 32\right\rangle
_S+0.00807\left| -\frac 72,\frac 52\right\rangle _S \\
&&+2.01\times 10^{-4}\left| -\frac 92,\frac 72\right\rangle _S-0.73604\left|
-\frac 12,-\frac 12\right\rangle ,
\end{eqnarray*}

\[
\left| \psi _{24}\right\rangle _A=0.68759\left| -\frac 32,\frac 12%
\right\rangle _A+0.16456\left| -\frac 52,\frac 32\right\rangle
_A+0.01192\left| -\frac 72,\frac 52\right\rangle _S+2.968\times
10^{-4}\left| -\frac 92,\frac 72\right\rangle _A, 
\]

\begin{eqnarray*}
\left| \psi _{25}\right\rangle _S &=&0.86948\left| -\frac 12,-\frac 12%
\right\rangle +0.34374\left| -\frac 32,\frac 12\right\rangle
_S+0.06184\left| -\frac 52,\frac 32\right\rangle _S \\
&&+0.00399\left| -\frac 72,\frac 52\right\rangle _S+8.347\times
10^{-5}\left| -\frac 92,\frac 72\right\rangle _S,
\end{eqnarray*}

\begin{eqnarray*}
\left| \psi _{26}\right\rangle _S &=&0.7067\left| -\frac 92,\frac 92%
\right\rangle _S-0.02399\left| -\frac 72,\frac 72\right\rangle _S+8.41\times
10^{-4}\left| -\frac 52,\frac 52\right\rangle _S \\
&&-3.02\times 10^{-5}\left| -\frac 32,\frac 32\right\rangle _S+9.43\times
10^{-7}\left| -\frac 12,\frac 12\right\rangle _S,
\end{eqnarray*}

\begin{eqnarray*}
\left| \psi _{26}\right\rangle _A &=&0.7067\left| -\frac 92,\frac 92%
\right\rangle _A-0.02399\left| -\frac 72,\frac 72\right\rangle _A+8.41\times
10^{-4}\left| -\frac 52,\frac 52\right\rangle _A \\
&&-3.02\times 10^{-5}\left| -\frac 32,\frac 32\right\rangle _A+1.1\times
10^{-6}\left| -\frac 12,\frac 12\right\rangle _A,
\end{eqnarray*}

\begin{eqnarray*}
\left| \psi _{27}\right\rangle _S &=&0.70427\left| -\frac 72,\frac 72%
\right\rangle _S+0.02391\left| -\frac 92,\frac 92\right\rangle
_S-0.05845\left| -\frac 52,\frac 52\right\rangle _S \\
&&+0.00385\left| -\frac 32,\frac 32\right\rangle _S-2.272\times
10^{-4}\left| -\frac 12,\frac 12\right\rangle _S,
\end{eqnarray*}

\begin{eqnarray*}
\left| \psi _{27}\right\rangle _A &=&0.70427\left| -\frac 72,\frac 72%
\right\rangle _A+0.02391\left| -\frac 92,\frac 92\right\rangle
_A-0.05845\left| -\frac 52,\frac 52\right\rangle _A \\
&&+0.00385\left| -\frac 32,\frac 32\right\rangle _A-2.56\times 10^{-4}\left|
-\frac 12,\frac 12\right\rangle _A,
\end{eqnarray*}

\begin{eqnarray*}
\left| \psi _{28}\right\rangle _S &=&0.69505\left| -\frac 52,\frac 52%
\right\rangle _S+0.05768\left| -\frac 72,\frac 72\right\rangle
_S+1.133\times 10^{-3}\left| -\frac 92,\frac 92\right\rangle _S \\
&&-0.11579\left| -\frac 32,\frac 32\right\rangle _S+0.01288\left| -\frac 12,%
\frac 12\right\rangle _S,
\end{eqnarray*}

\begin{eqnarray*}
\left| \psi _{28}\right\rangle _A &=&0.69485\left| -\frac 52,\frac 52%
\right\rangle _A+0.05767\left| -\frac 72,\frac 72\right\rangle
_A+0.00113\left| -\frac 92,\frac 92\right\rangle _A \\
&&-0.11651\left| -\frac 32,\frac 32\right\rangle _A+0.01684\left| -\frac 12,%
\frac 12\right\rangle _A,
\end{eqnarray*}

\begin{eqnarray*}
\left| \psi _{29}\right\rangle _S &=&0.67998\left| -\frac 32,\frac 32%
\right\rangle _S+0.11365\left| -\frac 52,\frac 52\right\rangle
_S+0.00571\left| -\frac 72,\frac 72\right\rangle _S \\
&&+8.81\times 10^{-5}\left| -\frac 92,\frac 92\right\rangle _S-0.15712\left|
-\frac 12,\frac 12\right\rangle _S,
\end{eqnarray*}

\begin{eqnarray*}
\left| \psi _{29}\right\rangle _A &=&0.61463\left| -\frac 32,\frac 32%
\right\rangle _A+0.10272\left| -\frac 52,\frac 52\right\rangle
_A+0.00516\left| -\frac 72,\frac 72\right\rangle _A \\
&&+7.96\times 10^{-5}\left| -\frac 92,\frac 92\right\rangle _A-0.33414\left|
-\frac 12,\frac 12\right\rangle _A,
\end{eqnarray*}

\begin{eqnarray*}
\left| \psi _{30}\right\rangle _{S,A} &=&0.65883\left| -\frac 12,\frac 12%
\right\rangle _{S,A}+0.2552\left| -\frac 32,\frac 32\right\rangle
_{S,A}+0.02857\left| -\frac 52,\frac 52\right\rangle _{S,A} \\
&&+1.202\times 10^{-3}\left| -\frac 72,\frac 72\right\rangle
_{S,A}+1.67\times 10^{-5}\left| -\frac 92,\frac 92\right\rangle _{S,A},
\end{eqnarray*}

\begin{eqnarray*}
\left| \psi _{31}\right\rangle _S &=&0.70614\left| \frac 92,-\frac 72%
\right\rangle _S-0.03701\left| \frac 72,-\frac 52\right\rangle _S+1.75\times
10^{-3}\left| \frac 52,-\frac 32\right\rangle _S \\
&&-8.163\times 10^{-5}\left| \frac 32,-\frac 12\right\rangle _S+7.77\times
10^{-6}\left| \frac 12,\frac 12\right\rangle ,
\end{eqnarray*}

\[
\left| \psi _{31}\right\rangle _A=0.70614\left| \frac 92,-\frac 72%
\right\rangle _A-0.03701\left| \frac 72,-\frac 52\right\rangle _S+1.75\times
10^{-3}\left| \frac 52,-\frac 32\right\rangle _A-8.163\times 10^{-5}\left| 
\frac 32,-\frac 12\right\rangle _A, 
\]

\begin{eqnarray*}
\left| \psi _{32}\right\rangle _S &=&0.70148\left| \frac 72,-\frac 52%
\right\rangle _S+0.03676\left| \frac 92,-\frac 72\right\rangle
_S-0.08077\left| \frac 52,-\frac 32\right\rangle _S \\
&&+0.00723\left| \frac 32,-\frac 12\right\rangle _S-1.236\times
10^{-3}\left| \frac 12,\frac 12\right\rangle ,
\end{eqnarray*}

\[
\left| \psi _{32}\right\rangle _A=0.70148\left| \frac 72,-\frac 52%
\right\rangle _S+0.03676\left| \frac 92,-\frac 72\right\rangle
_A-0.08075\left| \frac 52,-\frac 32\right\rangle _A+0.0071\left| \frac 32,-%
\frac 12\right\rangle _A, 
\]

\[
\left| \psi _{33}\right\rangle _S=0.70536\left| \frac 92,-\frac 52%
\right\rangle _S-0.04963\left| \frac 72,-\frac 32\right\rangle
_S+3.012\times 10^{-3}\left| \frac 52,-\frac 12\right\rangle _S-1.68\times
10^{-4}\left| \frac 32,\frac 12\right\rangle _S, 
\]

\[
\left| \psi _{33}\right\rangle _A=0.70536\left| \frac 92,-\frac 52%
\right\rangle _A-0.04963\left| \frac 72,-\frac 32\right\rangle
_A+3.012\times 10^{-3}\left| \frac 52,-\frac 12\right\rangle
_A-0.00019\left| \frac 32,\frac 12\right\rangle _A, 
\]

\[
\left| \psi _{34}\right\rangle _S=0.69716\left| \frac 72,-\frac 32%
\right\rangle _S+0.04905\left| \frac 92,-\frac 52\right\rangle
_S-0.10696\left| \frac 52,-\frac 12\right\rangle _S+0.01129\left| \frac 32,%
\frac 12\right\rangle _S, 
\]

\[
\left| \psi _{34}\right\rangle _A=0.69696\left| \frac 72,-\frac 32%
\right\rangle _A+0.04904\left| \frac 92,-\frac 52\right\rangle
_A-0.10782\left| \frac 52,-\frac 12\right\rangle _A+0.01455\left| \frac 32,%
\frac 12\right\rangle _A, 
\]

\[
\left| \psi _{35}\right\rangle _S=0.70425\left| \frac 92,-\frac 32%
\right\rangle _S-0.0633\left| \frac 72,-\frac 12\right\rangle
_S+0.0049\left| \frac 52,\frac 12\right\rangle _S-7.49\times 10^{-4}\left| 
\frac 32,\frac 32\right\rangle , 
\]

\[
\left| \psi _{35}\right\rangle _A=0.70425\left| \frac 92,-\frac 32%
\right\rangle _A-0.06328\left| \frac 72,-\frac 12\right\rangle
_A+0.00484\left| \frac 52,\frac 12\right\rangle _A, 
\]

\[
\left| \psi _{36}\right\rangle _S=0.7026\left| \frac 92,-\frac 12%
\right\rangle _S-0.07938\left| \frac 72,\frac 12\right\rangle
_S+0.00728\left| \frac 52,\frac 32\right\rangle _S, 
\]

\[
\left| \psi _{36}\right\rangle _A=0.70256\left| \frac 92,-\frac 12%
\right\rangle _A-0.07964\left| \frac 72,\frac 12\right\rangle
_A+0.00846\left| \frac 52,\frac 32\right\rangle _A. 
\]

{\bf References}

[1] Sessoli R, Gatteschi D, Caneschi A and Novak M A , Nature 365, (1993)141

[2] Aubin S M J and et al., J. Am. Chem. Soc. 118, (1996)7746

[3] Boskovic C and et al., J. Am. Chem. Soc. 124, (2002)3725

[4] Friedman J R , Sarachik M P , Tejada J and Ziolo R, Phys. Rev. Lett. 76,
(1996)3830

[5] Thomas L, Lionti F, Ballou R, Gatteschi D, Sessoli R and Barbara B,
Nature 383, (1996)145

[6] Sangregorio C, Ohm T, Paulsen C, Sessoli R and Gatteschi D, Phys. Rev.
Lett. 78, (1997)4645

[7] Hill S and et al., Phys. Rev. Lett. 80, (1998)2453

[8] Bokacheva L, Kent A D, and Walters M A, Phys. Rev. Lett. 85, (2000)4803

[9] Wernsdorfer W and Sessoli R, Science 284, (1999)133

[10] Garg A, EuroPhys. Lett. 22, (1993)205

[11] Caneschi A, Gatteschi D, Sangregorio C, Sessoli R, Sorace L, Cornia A,
Novak M A, Paulsen C, Wernsdorfer W, J. Magn. Magn. Mater. 200, (1999)182

[12] Liang J-Q, M$\ddot{u}$ller-Kirsten H J W, Park D K and Zimmerschied F,
Phys. Rev. Lett. 81, (1998)216

[13] Kou S P, Liang J-Q, Zhang Y-B, Wang X-B, Pu F-C, Phys. Rev. B59,
(1999)6309

[14] Liang J-Q, Zhang Y-B, M$\ddot{u}$ller-Kirsten H J W, Zhou J G,
Zimmerschied F and Pu F-C, Phys. Rev. B57, (1998)529

[15] Liang J-Q, M$\ddot{u}$ller-Kirsten H J W, Park D K and Pu F-C, Phys.
Rev. B61, (2000)8856

[16] Nie Yi-Hang, Jin Yan-Hong, Liang J-Q and Pu F-C, Phys. Rev. B64,
(2001)134417

[17] Zhou B, Liang J-Q and Pu F-C, Phys. Rev. B64, (2001)132407

[18] Leuenberger M N and Loss D, Nature 410, (2001)789

[19] Benjamin S C and Bose S, Phys. Rev. Lett. 90, (2003)247901

[20] Meier F, Levy J and Loss D, Phys. Rev. B68, (2003)134417

[21] Park K, Novotny M A, Dalal N S, Hill S,\ Rikvold P A, Phys. Rev. B65,
(2002)014426

[22] Hill S, Maccagnano S, Park K, Achey R M, North J M, Dalal N S, Phys.
Rev. B65, (2002)224410

[23] Park K, Novotny M A, Dalal N S, Hill S,\ Rikvold P A, Phys. Rev. B66,
(2002)144409

[24] Zipse D, North J M, Dalal N S, Hill S, Edwards R S, Phys. Rev. B68,
(2003)184408

[25] Wernsdorfer W, Aliaga-Alcalde N, Hendrickson D N and Christou G, Nature
416, (2002)406

[26] Wernsdorfer W, Aliaga-Alcalde N, Tiron R, Hendrickson D N and Christou
G, J. Magn. Magn. Mater. 272-276, (2004)1037

[27] Tiron R, Wernsdorfer W, Foguet-Albiol D, Aliaga-Alcalde N and Christou
G, Phys. Rev. Lett. 91, (2003)227203

[28] Wernsdorfer W, Bhaduri S, Tiron R, Hendrickson D N and Christou G,
Phys. Rev. Lett. 89, (2002)197201

[29] Kim Gwang-Hee, Phys. Rev. B67, (2003)024421

[30] Su Yuanchang and Tao Ruibao, Phys. Rev. B68, (2003)024431

[31] Hill S, Edwards R S, Aliaga-Alcalde N and Christou G, Science 302,
(2003)1015

[32] GaraninD A, J. Phys. A: Math. Gen. 24, (1991)L61

[33] Bennett C H, Bernstein H J, Popescu S, Schumacher B, Phys. Rev. A53,
(1996)2046

Table caption

Table 1. The resonant field transitions. Theoretical values are obtained by
the high-order level splitting formula and eigenvectors with parameter
values $D=0.72$K, $B=1.8\times 10^{-3}$K and $J=0.10$K.

Table 2. Level splitting $\triangle E$ in Eq. ($22$).

Figure caption

Fig. 1. Schematic display of lowest energy states (M$=-9$ to $-4$) and the
part of lower energy states (M$=-3$ to $1$). The zeroth-order energy levels
and eigenvectors are shown on the left; energy shifts due to the
exchange-interaction are shown in the center; and the results of the
high-order perturbation calculation are displayed on the right. The red
lines[x, (a) to (i)] denote the strongest EPR\ transitions; the blue lines
denote the weaker EPR\ transitions; the magenta lines denote other possible
EPR\ transitions.

Fig. 2. EPR- peak positions ( solid lines) obtained from Eq. ($23$) with
different frequencies.

\end{document}